\documentclass{ifacconf}

\usepackage{graphicx}      
\usepackage{natbib}        
\usepackage{xcolor}%
\usepackage{amsmath}%
\usepackage{tikz}%
\usepackage{siunitx}%
\usepackage{import}
\usepackage[ruled,vlined,algo2e]{algorithm2e}%

\begin{document}
\begin{frontmatter}

\title{Friction and Road Condition Estimation by Combining Cause- and Effect-Based Methods using Bayesian Networks}


\author[First]{Bj{\"o}rn Volkmann} 
\author[First]{Karl-Philipp Kortmann} 
\author[Second]{Ulrich Mair} 
\author[Second]{Julian King} 

\address[First]{Leibniz University Hannover, Institute of Mechatronic Systems, \\
   An der Universit{\"a}t 1, 30823 Garbsen, Germany (e-mail: bjoern.volkmann@imes.uni-hannover.de)}
\address[Second]{ZF Friedrichshafen AG, \\
Graf-von-Soden-Platz 1, 88046 Friedrichshafen, Germany}

\begin{abstract}                
Knowledge about the maximum tire-road friction potential is an important factor to ensure the driving stability and traffic safety of the vehicle. Many authors proposed systems that either measure friction related parameters or estimate the friction coefficient directly via a mathematical model. However these systems can be negatively impacted by environmental factors or require a sufficient level of excitation in the form of tire slip, which is often too low under practical conditions. Therefore, this work investigates, if a more robust estimation can be achieved by fusing the information of multiple systems using a Bayesian network, which models the statistical relationship between the sensors and the maximum friction coefficient. First, the Bayesian network is evaluated over its entire domain to compare the inference process to all possible road conditions. After that, the algorithm is applied to data from a test vehicle to demonstrate the performance under real conditions.
\end{abstract}


\end{frontmatter}

\section{Introduction} \label{ch:introduction}
The development of advanced driver assistance systems like antilock braking system (ABS), acceleration slip regulation (ASR), electronic stability control (ESC) or collision avoidance system (CAS) has contributed to the decline of fatal road accidents over the last decades. For these systems, information on the current road condition and tire-road friction is an important factor to ensure the driving stability and traffic safety of the vehicle. However, the tire-road friction is not directly measured in current commercially available vehicles because the required sensors would be too expensive. Instead, an estimation approach is required.

According to \cite{Muller2003} the algorithms to estimate parameters related to the tire-road friction can be categorized into the two groups of \textit{cause-based} and \textit{effect-based} methods. \cite{Khaleghian2017} and \cite{Acosta2017} provide a review on different examples for these approaches.

Cause-based methods correlate sensor data to parameters which influence tire-road friction to give an estimation of the maximum friction potential. An advantage of such methods is their ability provide information on the friction potential, even during situations of low excitation like free-rolling driving conditions. However, according to \cite{Khaleghian2017} and \cite{Muller2003}, the confidence and accuracy of the estimation can degrade with environmental factors that were not considered in the training data. Examples for cause-based methods are the classification of the road surface using camera images by \cite{Nolte2018}, \cite{Busch2020} and \cite{Fink2020} or the estimation of road roughness using acoustic data by \cite{Gabrielli2019}.

Effect-based methods measure the effects of the tire-friction to make an estimation of the maximum friction potential. These can be further categorized into acoustic, tire-tread and slip-based approaches. According to \cite{Acosta2017}, especially slip-based approaches have received much attention in the literature since the required sensors are already available on modern vehicles. However, as mentioned in \cite{Kiencke2005} and \cite{Muller2003}, slip-based approaches generally require a sufficient level of slip in the tire-road contact, with the friction potential close to its maximum, while in most driving situations the experienced slip is relatively small. For slip-based methods the authors \cite{Rajamani2012},  \cite{Zhao2014} and \cite{Singh2015} have proposed systems that estimate tire forces and slip using an observer and then identifying a tire-model with a recursive least squares algorithm. There are also Bayesian approaches by \cite{Berntorp2020}, where the the relationship between slip and tire-forces is identified using Gaussian processes. Another approach by \cite{Wielitzka2018} is the estimation of the tire-road friction coefficient directly with an observer that uses a model of the longitudinal- and lateral vehicle dynamics.

With the given advantages and disadvantages of the above mentioned approaches, it may be practical to consider methods of data fusion to combine strengths and compensate weaknesses of the different systems, with the goal to generate a more robust estimation of the tire-road friction. \cite{Jonsson2011} fuses camera images with weather information to classify the weather condition of the road surface. While it was shown that additional information about environmental conditions could enhance the result of the classification, the labels of the proposed model did not contain information on the pavement type. Additionally, knowledge about the road surface condition can only provide coarse information on the maximum tire friction potential. The results could be further enhanced by effect-based methods during situations of sufficient slip. \cite{Leng2022} use camera images to increase the robustness of a dynamics based friction estimator. The information from the camera is used to better identify situations of sufficient slip in the tire-road contact. However, the estimator is still reliant on sufficient slip to enable an estimation of the friction potential.

Given the possible advantages of data fusion, this paper investigates if a more robust estimation of the tire-friction potential can be achieved by fusing the information generated by multiple cause- and effect-based estimators using a Bayesian network (BN). The goal of this approach is to compensate weaknesses of individual sources of information and increase robustness of the estimation due to the available information from multiple sources. The road condition is represented in form of a probability distribution. In this context a higher robustness means providing a confidence interval which is as small as possible, while keeping its probability as high as possible. Applying an Bayesian network for the estimation of the tire-road friction coefficient was first proposed in a patent by \cite{Schlegel2017}. 

The considered sensors are an RGB-camera, acoustic sensors, the outside-air temperature sensor and a vehicle-dynamics based friction observer.
Through the use of a BN it is possible to take statistical relations between the maximum friction potential and related parameters like pavement type or surface condition into account. The BN proposed in this paper only contains discrete probability distributions. The information generated by the above mentioned systems is used as evidence in the BN to infer the road condition and the tire-road friction potential.
The Bayesian network is evaluated over its entire domain  to compare the results of an inference with given evidence to all possible road conditions. Additionally, the algorithm is applied to data from a test vehicle to evaluate its performance under real conditions.

\section{Sources of Information} \label{ch:sensors}
\begin{figure}[h]
\begin{center}
\begingroup%
  \makeatletter%
  \providecommand\color[2][]{%
    \errmessage{(Inkscape) Color is used for the text in Inkscape, but the package 'color.sty' is not loaded}%
    \renewcommand\color[2][]{}%
  }%
  \providecommand\transparent[1]{%
    \errmessage{(Inkscape) Transparency is used (non-zero) for the text in Inkscape, but the package 'transparent.sty' is not loaded}%
    \renewcommand\transparent[1]{}%
  }%
  \providecommand\rotatebox[2]{#2}%
  \newcommand*\fsize{\dimexpr\f@size pt\relax}%
  \newcommand*\lineheight[1]{\fontsize{\fsize}{#1\fsize}\selectfont}%
  \ifx\svgwidth\undefined%
    \setlength{\unitlength}{239.54697798bp}%
    \ifx\svgscale\undefined%
      \relax%
    \else%
      \setlength{\unitlength}{\unitlength * \real{\svgscale}}%
    \fi%
  \else%
    \setlength{\unitlength}{\svgwidth}%
  \fi%
  \global\let\svgwidth\undefined%
  \global\let\svgscale\undefined%
  \makeatother%
  \begin{picture}(1,0.7751662)%
    \lineheight{1}%
    \setlength\tabcolsep{0pt}%
    \put(0,0){\includegraphics[width=\unitlength,page=1]{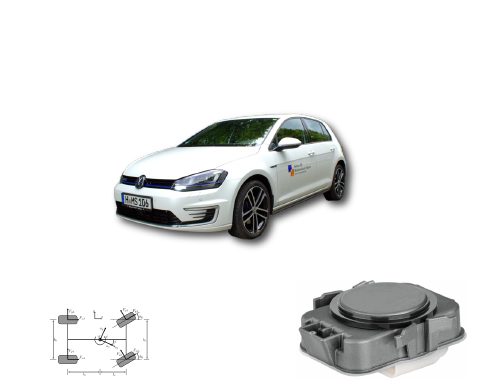}}%
    \put(0.19839351,0.23303124){\makebox(0,0)[t]{\lineheight{1.25}\smash{\begin{tabular}[t]{c}Model-Based\\Friction Estimator\end{tabular}}}}%
    \put(0.75786796,0.22952634){\makebox(0,0)[t]{\lineheight{1.25}\smash{\begin{tabular}[t]{c}Piezo-Acoustic Sensor\end{tabular}}}}%
    \put(0.11950525,0.72020557){\makebox(0,0)[t]{\lineheight{1.25}\smash{\begin{tabular}[t]{c}Camera\end{tabular}}}}%
    \put(0.78488807,0.72162667){\makebox(0,0)[t]{\lineheight{1.25}\smash{\begin{tabular}[t]{c}Temperature Sensor\end{tabular}}}}%
    \put(0,0){\includegraphics[width=\unitlength,page=2]{SensorSetup.pdf}}%
  \end{picture}%
\endgroup%

	\caption{Vehicle equipped with different sensor systems and a model based friction estimator}
	\label{fig:SensorSetup}
\end{center}
\end{figure}

The considered sensors are a camera system for the optical classification of the road surface condition, two piezo-acoustic sensors to detect the road weather condition, the serial installed temperature sensor and a model-based friction estimator which considers combined longitudinal and lateral vehicle dynamics. The sensor setup is shown in Fig.\,\ref{fig:SensorSetup}.

\subsection{Camera System} \label{ch:camera}
The purpose of this sensor is the multi-class classification of the road surface condition in terms of pavement type and weather condition. For tasks like image classification, \cite{Nolte2018}, \cite{Busch2020} and \cite{Fink2020} have shown that the usage of deep convolutional neural networks are an effective solution which can achieve high accuracies. In this paper, a pre-trained SqueezeNet-architecture from the \textit{MATLAB Deep-Learning Toolbox} is used for image classification. The data set used for training is a compilation of publicly available data sets (KITTI from \citealp{Geiger2013IJRR}, RobotCar from \citealp{RobotCarDatasetIJRR}, CityScapes from \citealp{Cordts2016Cityscapes}, BDD100K from \citealp{Yu2018}) as well as self recorded images. Only a rectangular region of interest of the road in front of the vehicle is used for the classification. The SqueezeNet has to differentiate the classes
\begin{itemize}
	\item Asphalt Dry (AD),
	\item Asphalt Wet (AW),
	\item Concrete Dry (CD),
	\item Concrete Wet (CW),
	\item Cobblestone Dry (CbD),
	\item Cobblestone Wet (CbW),
	\item and Snow (S).
\end{itemize}

\begin{figure}[h]
\begin{center}
	\resizebox{8.4cm}{!}{\import{Figures/ReproBilder/}{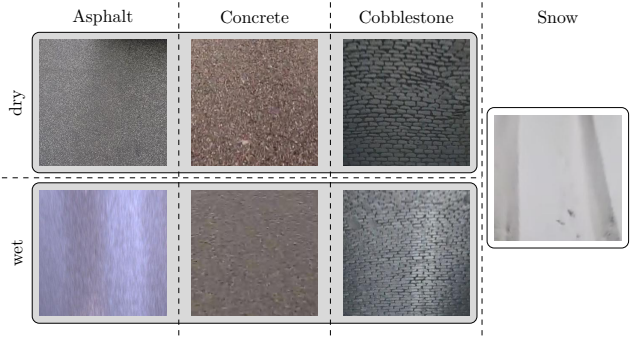}}
	\caption{Exemplary pictures from the dataset}
	\label{fig:ReproBilder}
\end{center}
\end{figure}

These classes contain no combination for \textit{snow} and pavement type, since a layer of snow would block the view on the underlying pavement. Examples of the images for every class are shown in Fig.\,\ref{fig:ReproBilder}.

\begin{figure}[h]
\begin{center}
	\includegraphics{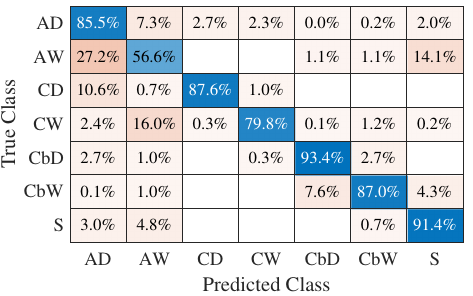}
	\caption{Confusion matrix for the SqueezeNet}
	\label{fig:SqueezeNet_CMAT}
\end{center}
\end{figure}

The confusion matrix is shown in Fig.\,\ref{fig:SqueezeNet_CMAT} and was generated from a test set of images not included in the training data. During the image classification the performance of the SqueezeNet can be negatively impacted by environmental factors like lighting, view angle and visibility through the windshield.

 \subsection{Piezo-Acoustic Sensor} \label{ch:RCS}
The piezo-acoustic road condition sensor (RCS) is manufactured by \textit{HELLA} and mounted in the wheel house behind the two front wheels of the vehicle. The sensor uses a piezo-electric element to detect drops of water that are whirled by the rotation of the wheels. Information is provided as a level of wetness which is a \SI{16}{bit} integer and used to infer the weather condition on the road surface. The sensor also has physical limitations. Detection of water is only possible if drops of water can impact on the piezo-electric surface. For this reason, the detection ability depends on the amount of water present on the road-surface and the surface's geometry. It is also not possible to differentiate water and snow.
 
 \subsection{Model-Based Friction Estimator}
The model-based friction observer is implemented as an unscented Kalman-filter. The system uses the dynamics of a two-track model to estimate vehicle states and the tire-friction coefficient. The tire-road contact forces are modeled with the magic-tire formula proposed by \cite{Pacejka2012}. The state vector

\begin{equation}
	\boldsymbol{x} = (\dot{\psi}, \beta, \omega_1, \omega_2, \omega_3, \omega_4, v, \mu_\mathrm{max} )^\mathrm{T}
\end{equation}

of the vehicle contains the yaw-rate $\dot{\psi}$, side-slip angle $\beta$, wheel rates $\omega_i$ and vehicle velocity $v$. For the parameter estimation the state vector is also augmented by the tire-friction coefficient $\mu_\mathrm{max}$. A similar system was proposed by \cite{Wielitzka2018} and \cite{Antonov2011}.

As mentioned in Sec.\,\ref{ch:introduction}, the model-based estimation of the tire-friction coefficient is only possible during phases of sufficient excitation. For the $i$-th tire the level of excitation is monitored using the sensitivity $S_{\mu,x,i}$ and $S_{\mu,y,i}$ of the longitudinal and lateral tire forces with respect to the friction coefficient. These values are calculated with

\begin{equation} \label{eq:sens_mu}
	S_{\mu,x,i} = \frac{\delta F_{x,i}}{\delta \mu_\mathrm{max}} \frac{1}{F_{z,i}}.
\end{equation}

Where $F_{x,i}$ are the contact forces in the tire's longitudinal direction. The sensitivity $S_{\mu,y,i}$ is calculated with equation\,\ref{eq:sens_mu} using the lateral tire forces $F_{y,i}$. Here $F_{z,i}$ is the respective vertical tire force. With equation\,\ref{eq:sens_mu} the sensitivity $S_{\mu,x,i}$ and $S_{\mu,y,i}$ describe the relative change of the longitudinal and lateral tire force with respect to the maximum friction coefficient. The estimation of the friction coefficient is only enabled when a sensitivity value is above a set threshold. During phases of low excitation the model-based friction estimator provides no evidence to the Bayesian network.

\subsection{Temperature Sensor}
The temperature sensor is a standard sensor that measures the outside air-temperature and can be found on almost all modern vehicles. Information on the surrounding air-temperature is used to make an assumption about current weather conditions. For the weather conditions considered in this paper, the temperature information is used to infer if water on the road surface is more likely to be liquid or frozen. 

Because only measurements of air temperature were available, the air-to-pavement temperature model, which will be presented in Sec\,\ref{ch:DFM}, is relatively simple. However, there have been efforts to also include data from Road Weather Information Systems (RWIS) to enhance predictions of the road surface temperature by \cite{Almkvist2023}. While the aforementioned model was not available for the results presented in this work, the inclusion of such advanced weather models could further improve the distinction of different road weather conditions.

\section{Data Fusion Model} \label{ch:DFM}
This paper investigates the estimation of the tire-friction potential by fusing information from multiple sources using a Bayesian network. A Bayesian network is a directed acyclic graph (DAG) which represents a probability distribution $\mathcal{P}(\boldsymbol{X}_1, \ldots, \boldsymbol{X}_n)$ over the domain $\boldsymbol{D} = \{\boldsymbol{X}_1, \ldots, \boldsymbol{X}_n\}$ with $n$ random variables $\boldsymbol{X}_i$. Each node in the DAG is associated with one variable on the domain and contains the conditional probability distribution $\mathcal{P}(\boldsymbol{X}_i \mid \mathrm{Parents}(\boldsymbol{X}_i))$ for the variable $\boldsymbol{X}_i$ given its parents $\mathrm{Parents}(\boldsymbol{X}_i)$. The joint probability distribution over the domain can then be calculated with

\begin{equation}
	\mathcal{P}(\boldsymbol{X}_1, \ldots, \boldsymbol{X}_n) = \prod_{i=1}^n \mathcal{P}(\boldsymbol{X}_i \mid \mathrm{Parents}(\boldsymbol{X}_i)).
\end{equation}

Given a set of evidence $\boldsymbol{E} \subseteq \boldsymbol{D}$ it is possible to calculate the conditional probability distribution $\mathcal{P}(\boldsymbol{Y} \mid \boldsymbol{E})$ for the variables $\boldsymbol{Y} \subseteq \boldsymbol{D} \setminus \boldsymbol{E}$ according to

\begin{equation}
	\mathcal{P}(\boldsymbol{Y} \mid \boldsymbol{E}) = \alpha \sum_{\boldsymbol{H}} \mathcal{P}(\boldsymbol{Y} , \boldsymbol{H}, \boldsymbol{E}).
	\label{eq:BN_query}
\end{equation}

Here $\boldsymbol{H}$ are the remaining variables of the domain with $\boldsymbol{H} = \boldsymbol{D} \setminus \{\boldsymbol{Y}, \boldsymbol{E}\}$ and $\alpha$ is a normalizing factor. In this paper the DAG represents a discrete probability distribution and the sensors described in Sec.~\ref{ch:sensors} will provide the evidence to infer the road condition. More information on the definition of Bayesian networks can be found in \cite{Russell2021} or \cite{Gaag2014}. 

\subsection{Network Topology}
\begin{figure}[h]
\begin{center}
	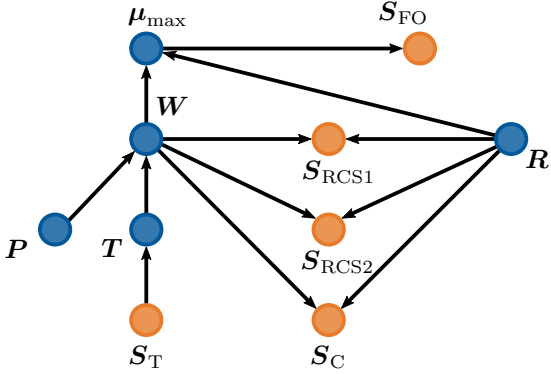
	\caption{Topology of the Bayesian network with nodes representing the road conditions (blue) and sensors (orange)}
	\label{fig:BN_Topology}
\end{center}
\end{figure}

The Bayesian network proposed in this paper represents a discrete probability distribution which correlates variables related to the tire-road friction to the available sensor information. The topology of the network can be seen in Fig.\,\ref{fig:BN_Topology}. The description of the nodes is given in Table\,\ref{tb:nodes}.

\begin{table}[h]
\begin{center}
\caption{Nodes of the Bayesian network}\label{tb:nodes}
\begin{tabular}{ccl}
	Index & Variable & Name \\\hline
	1 & $\boldsymbol{R}$ & Road pavement type \\
	2 & $\boldsymbol{T}$ & Pavement temperature \\ 
	3 & $\boldsymbol{P}$ & Precipitation \\
	4 & $\boldsymbol{W}$ & Road surface weather condition \\
	5 & $\boldsymbol{\mu}_\mathrm{max}$ & Maximum Road friction potential \\
	6 & $\boldsymbol{S}_\mathrm{C}$ & Camera \\
	7 & $\boldsymbol{S}_\mathrm{T}$ & Air-temperature sensor \\
	8 & $\boldsymbol{S}_\mathrm{RCS1}$ & Piezo-acoustic road condition sensor \\
	9 & $\boldsymbol{S}_\mathrm{RCS2}$ & Piezo-acoustic road condition sensor\\
	10 & $\boldsymbol{S}_\mathrm{FO}$ & Model based friction observer \\\hline
\end{tabular}
\end{center}
\end{table}

The central node of the BN is the maximum tire-road friction coefficient $\boldsymbol{\mu}_{\mathrm{max}}$. This variable is observed by the model based friction estimator $\boldsymbol{S}_{\mathrm{FO}}$ which is a child node of $\boldsymbol{\mu}_{\mathrm{max}}$. The friction potential is dependent on the surface type and potential lubricant between tire and road surface. These are the nodes for the pavement type $\boldsymbol{R}$ and the road weather condition $\boldsymbol{W}$. The road weather condition is observed by the RCS, introduced in Sec.\,\ref{ch:RCS}. The vehicle is equipped with two RCS which are mounted behind each of the front two tires respectively. Therefore, the BN contains one node for each RCS, which are $\boldsymbol{S}_\mathrm{RCS1}$ and $\boldsymbol{S}_\mathrm{RCS2}$ respectively. As mentioned in Sec.\,\ref{ch:camera} the labels of the camera system describe a combination of pavement type and weather condition. As a result the node for the camera $\boldsymbol{S}_\mathrm{C}$ is a child of both $\boldsymbol{R}$ and $\boldsymbol{W}$. The road weather condition $\boldsymbol{W}$ is dependent on precipitation $\boldsymbol{P}$ and pavement temperature $\boldsymbol{T}$. Generally, a node representing a sensor would be the child node of the phenomenon which is measured. In the case of air temperature, changes are expected to happen over a longer period of time and the temperature sensor is assumed to be quasi deterministic with respect to the true air temperature. Therefore, the air temperature is directly represented as the node of the temperature sensor $\boldsymbol{S}_\mathrm{T}$ and is the parent of the node representing the pavement temperature $\boldsymbol{T}$.

\subsection{Model Parameters}
The parameters of the BN are the entries of the conditional probability distributions for its nodes. Since the probability distribution of the BN is discrete, all physical values must be presented as discrete classes. Due to the definition in terms of conditional probability distributions, it is possible to parametrise the BN from incremental sets of data that only contain values for a node and all its parents. If no data is available it is necessary to rely on expert judgment. For this purpose \cite{Verzobio2021} proposed a probability scale for a case study on dam failure, which was used as reference in this paper. 

The root nodes describe environmental conditions whose states are constant over a certain period of time. Initial test showed that distributions in the root nodes which lean towards a specific class will negatively impact the classification of the other classes. This also affects their descendants.  In the case of camera and pavement type, a correct classification from the camera will likely be interpreted as an incorrect classification for the pavement type with the highest probability. This means, that the inference would be less accurate during environmental conditions with a lower probability. Therefore, the root nodes of the Bayesian network proposed here are represented by a uniform distribution.

\subsubsection{Pavement Type:}
The pavement types considered in this paper are \textit{asphalt}, \textit{concrete} and \textit{cobblestone}. The pavement type is a root node and is quantified as a uniform distribution.

\subsubsection{Air Temperature:}
The air temperature is discretized into the following classes:
\begin{itemize}
	\item $S_{T,1}$ : $S_T$ $>$ \SI{5}{\celsius}
	\item $S_{T,2}$ : \SI{5}{\celsius} $\geq$ $S_T$ $>$ \SI{0}{\celsius}
	\item $S_{T,3}$ : \SI{0}{\celsius} $\geq$ $S_T$ $>$ \SI{-21}{\celsius}
	\item $S_{T,4}$ : \SI{-21}{\celsius} $\geq$ $S_T$
\end{itemize}
This discretization was chosen under consideration of the dew temperature of pure water and a saturated saline solution. Since the air temperature $\boldsymbol{S}_\mathrm{T} \in \boldsymbol{E}$ is a root node, the probability distribution $\mathcal{P}(\boldsymbol{S}_\mathrm{T})$ becomes a scalar value in equation~\ref{eq:BN_query} and is eliminated with the normalization by $\alpha$. Therefore, the distribution of $\boldsymbol{S}_\mathrm{T}$ can be chosen as uniform.

\subsubsection{Pavement Temperature:}
The pavement temperature is discretized in the same intervals as the air temperature. The correlation between air and pavement temperature was calculated with a road-weather-information-system (RWIS) data set from the \textit{Iowa Environmental Mesonet} of the \textit{Iowa State University}. The used data is public domain and originates from multiple weather stations in 30 US-States. The used data ranges from 2018 to 2020 and contains roughly 980 million samples of air and pavement temperature. The resulting distribution is shown in Table\,\ref{tb:PD_PT}.
\begin{table}[h]
\begin{center}
\caption{Probability distribution for the pavement temperature $\boldsymbol{T}$}\label{tb:PD_PT}
\begin{tabular}{lrrrr}
	Air Temperature & $T_1$ & $T_2$ & $T_3$ & $T_4$ \\\hline
	$S_{T,1}$ & 95.05 \% & 1.84 \% & 0.87 \% & 0.24 \% \\
	$S_{T,2}$ & 41.46 \% & 50.73 \% & 7.54 \% & 0.27 \% \\
	$S_{T,3}$ & 5.07 \% & 22.68 \% & 71.72 \% & 0.53 \% \\
	$S_{T,4}$ & 10.15 \% & 2.87 \% & 51.40 \% & 35.58 \% \\\hline
\end{tabular}
\end{center}
\end{table}

\subsubsection{Precipitation:}
The precipitation is classified as \textit{true} or \textit{false}. The precipitation is a root node and is quantified as a uniform distribution.

\subsubsection{Road Weather Condition:}
For the road weather condition no data set was available. Therefore, the probability distribution has to be constructed manually. The resulting distribution is shown in Table\,\ref{tb:PD_RWC}.
\begin{table}[h]
\begin{center}
\caption{Probability distribution for the road weather condition $\boldsymbol{W}$}\label{tb:PD_RWC}
\begin{tabular}{llrrr}
	Precipitation & Pavement & Dry & Wet & Snow \\\hline
	True & $T_1$		& 5.00 \% & 95.00 \% & 0.00 \% \\
	True & $T_2$		& 5.00 \% & 90.00 \% & 5.00 \% \\
	True & $T_3$		& 5.00 \% & 20.00 \% & 75.00 \% \\
	True & $T_4$		& 5.00 \% & 0.00 \% & 95.00 \% \\
	False & $T_1$		& 95.00 \% & 5.00 \% & 0.00 \% \\
	False & $T_2$		& 95.00 \% & 2.50 \% & 2.50 \% \\
	False & $T_3$		& 95.00 \% & 1.75 \% & 3.25 \% \\
	False & $T_4$		& 95.00 \% & 0.00 \% & 5.00 \% \\\hline
\end{tabular}
\end{center}
\end{table}

\subsubsection{Maximum Friction Coefficient:}
The conditional probability distribution of the maximum friction coefficient was calculated with a data set provided by \cite{Muller2019}. The data set contains 3605 measurements of the friction coefficient on asphalt, concrete and cobblestone during different road weather conditions. The friction coefficient is discretised into 8 classes in the range of 0\,-\,1.2 with increments of 0.15. These classes will be called $\mu_{\mathrm{max},1}$ to $\mu_{\mathrm{max},8}$. The resulting distribution is shown in Table\,\ref{tb:PD_mumax}. In addition to road weather condition and pavement type, the maximum friction coefficient is significantly affected by the tire operating conditions in terms of inflation pressure, tread-depth, load, and tire temperature. These influences can be captured by adaptive tire models as presented in \cite{Singh2015a} but were beyond the scope of the current study.
\begin{table*}[h]
\begin{center}
\caption{Probability distribution for the maximum road friction $\boldsymbol{\mu}_\mathrm{max}$}\label{tb:PD_mumax}
\begin{tabular}{llrrrrrrrr}
	Pavement & Weather & $\mu_{\mathrm{max},1}$ & $\mu_{\mathrm{max},2}$ & $\mu_{\mathrm{max},3}$ & $\mu_{\mathrm{max},4}$ & $\mu_{\mathrm{max},5}$ & $\mu_{\mathrm{max},6}$ & $\mu_{\mathrm{max},7}$ & $\mu_{\mathrm{max},8}$ \\\hline
	Asphalt & Dry			& 0 \% & 0 \% & 0 \% & 0 \% & 0 \% & 15 \% & 76 \% & 9 \% \\
	Asphalt & Wet			& 0 \% & 0 \% & 0 \% & 11 \% & 47 \% & 36 \% & 5 \% & 0 \% \\
	Asphalt & Snow		& 7 \% & 51 \% & 3 \% & 9 \% & 2 \% & 1 \% & 0 \% & 0 \% \\
	Concrete & Dry		& 0 \% & 0 \% & 0 \% & 0 \% & 0 \% & 7 \% & 72 \% & 21 \% \\
	Concrete & Wet		& 0 \% & 0 \% & 0 \% & 0 \% & 7 \% & 87 \% & 6 \% & 0 \% \\
	Concrete & Snow		& 13 \% & 42 \% & 26 \% & 11 \% & 5 \% & 2 \% &1 \% & 0 \% \\
	Cobblestone & Dry		& 0 \% & 0 \% & 0 \% & 3 \% & 54 \% & 42 \% & 1 \% & 0 \% \\
	Cobblestone & Wet		& 0 \% & 9 \% & 72 \% & 18 \% & 1 \% & 0 \% & 0 \% & 0 \% \\
	Cobblestone & Snow	& 8 \% & 73 \% & 18 \% & 1 \% & 0 \% & 0 \% & 0 \% & 0 \% \\\hline
\end{tabular}
\end{center}
\end{table*}

\subsubsection{Model-Based Friction Observer:}
The conditional probability distribution of the friction observer is a challenging problem, since during driving tests, a ground truth of the road friction is generally not available. Therefore, the distribution was derived from simulations during situations with sufficient excitation. The distribution is shown Table\,\ref{tb:PD_DSO}.
\begin{table*}[h]
\begin{center}
\caption{Probability distribution for the model-based friction observer $\boldsymbol{S}_\mathrm{FO}$}\label{tb:PD_DSO}
\begin{tabular}{lrrrrrrrr}
	Friction Coefficient & $S_{\mathrm{FO},1}$ & $S_{\mathrm{FO},2}$ & $S_{\mathrm{FO},3}$ & $S_{\mathrm{FO},4}$ & $S_{\mathrm{FO},5}$ & $S_{\mathrm{FO},6}$ & $S_{\mathrm{FO},7}$ & $S_{\mathrm{FO},8}$ \\\hline
	$\mu_{\mathrm{max},1}$	& 99.68 \% & 0.02 \% & 0.00 \% & 0.00 \% & 0.27 \% & 0.00 \% & 0.03 \% & 0.00 \% \\
	$\mu_{\mathrm{max},2}$	& 71.76 \% & 20.7 \% & 0.21 \% & 0.01 \% & 6.62 \% & 0.00 \% & 0.70 \% & 0.00 \% \\
	$\mu_{\mathrm{max},3}$	& 55.76 \% & 0.15 \% & 28.37 \% & 9.55 \% & 5.25 \% & 0.00 \% & 0.92 \% & 0.00 \% \\
	$\mu_{\mathrm{max},4}$	& 17.31 \% & 0.04 \% & 0.1 \% & 56.31 \% & 22.96 \% & 2.85 \% & 0.43 \% & 0.00 \% \\
	$\mu_{\mathrm{max},5}$	& 10.70 \% & 0.02 \% & 0.03 \% & 0.54 \% & 78.5 \% & 8.91 \% & 1.09 \% & 0.21 \% \\
	$\mu_{\mathrm{max},6}$	& 5.32 \% & 0.01 \% & 0.02 \% & 0.21 \% & 10.47 \% & 75.01 \% & 8.79 \% & 0.17 \% \\
	$\mu_{\mathrm{max},7}$	& 2.12 \% & 0.01 \% & 0.01 \% & 0.05 \% & 8.03 \% & 7.84 \% & 81.49 \% & 0.45 \% \\
	$\mu_{\mathrm{max},8}$	& 0.45 \% & 0.02 \% & 0.02 \% & 0.03 \% & 2.24 \% & 2.59 \% & 19.12 \% & 75.53 \% \\\hline
\end{tabular}
\end{center}
\end{table*}

\subsubsection{Camera:}
The camera system was introduced in Sec.\,\ref{ch:camera}. The classes of the camera are a combination of pavement type and road weather condition. The conditional probability distributions are defined by the confusion matrix in Fig.\,\ref{fig:SqueezeNet_CMAT}. Combinations of pavement type with snow get the same probability distribution for the \textit{snow} label in the confusion matrix.

\subsubsection{Road Condition Sensor:}
The RCS classifies the road weather condition as level of wetness. The sensor signal is discretized into 3 classes. The distribution was summarized, adapted and quantified in cooperation with the manufacturer, based on data recorded with the available test vehicle. The probability distribution is shown in Table\,\ref{tb:PD_RCS}.
\begin{table}[h]
\begin{center}
\caption{Probability distribution for the road condition sensor $\boldsymbol{S}_\mathrm{RCS}$}\label{tb:PD_RCS}
\begin{tabular}{llrrr}
	$\boldsymbol{W}_1$ & $\boldsymbol{R}_1$ & $\textbf{S}_{\mathrm{RCS},1,1}$ & $\textbf{S}_{\mathrm{RCS},1,2}$ & $\textbf{S}_{\mathrm{RCS},1,3}$ \\\hline
	Dry & Asphalt $\vee$ Concrete & 95.00 \% & 5.00 \% & 0.00 \% \\
	Wet & Asphalt $\vee$ Concrete & 17.5 \% & 26.25 \% & 56.25 \% \\
	Snow & Asphalt $\vee$ Concrete & 96.00 \% & 4.00 \% & 0.00 \% \\
	Dry & Cobblestone & 99.00 \% & 1.00 \% & 0.00 \% \\
	Wet & Cobblestone & 99.00 \% & 1.00 \% & 0.00 \% \\
	Snow & Cobblestone & 99.00 \% & 1.00 \% & 0.00 \% \\\hline
\end{tabular}
\end{center}
\end{table}

\section{Model Evaluation}
With the Bayesian network described in Sec.\,\ref{ch:DFM}, it can now be quantified how the use of the considered sensors would benefit the estimation of the road condition. The domain $\boldsymbol{D} = \{\boldsymbol{X}, \boldsymbol{E}\}$ of the Bayesian network contains random variables representing the road conditions $\boldsymbol{X}$ and sensors $\boldsymbol{E}$ (Compare Fig.\,\ref{fig:BN_Topology} and Table\,\ref{tb:nodes}). Since all random variables are discrete there are $n$ possible combinations $\boldsymbol{d}_k = \{\boldsymbol{x}_k, \boldsymbol{e}_k\}$ for their values, whose probability is greater than zero. For all combinations $\boldsymbol{d}_k$ the probability distributions of all road conditions is calculated from the BN according to equation\,\ref{eq:BN_query}. Additionally, to show the influence of individual sensors, this evaluation is done for all combinations of the available sensors $\hat{\boldsymbol{E}} \in 2^{\boldsymbol{E}}$. The result of this process are the probability distributions $\mathcal{P}_k(\boldsymbol{X}_i \mid \hat{\boldsymbol{e}}_k)$ for the road condition $\boldsymbol{X}_i$ with the evidence $\hat{\boldsymbol{e}}_k$ in combination $k$. For further notation the estimate of variable $\boldsymbol{X}_i$ for combination $k$ will be called $\hat{\boldsymbol{X}}_{i,k}$ and the respective ground truth derived from $\boldsymbol{x}_k$ will be called $\boldsymbol{X}_{i,k}$. The goodness of fit for the distributions $\hat{\boldsymbol{X}}_{i,k}$ is quantified with error metrics for probability distributions. The values for road-friction coefficient are on a ratio scale and use the average Wasserstein-distance over all samples calculated by 

\begin{equation}\label{eq:avr_wass_dist}
	\overline{\mathcal{W}}_1(\boldsymbol{X}_i) = \frac{1}{n}\sum_{k=1}^n \mathcal{W}_1(\boldsymbol{X}_{i,k}, \hat{\boldsymbol{X}}_{i,k})  \text{, where}
\end{equation}

\begin{equation}\label{eq:wass_dist}
	\mathcal{W}_1(\boldsymbol{P}, \boldsymbol{Q}) = \sum_{x\in\boldsymbol{X}} |\mathcal{F}_P(\boldsymbol{X}) -\mathcal{F}_Q(\boldsymbol{X})|.
\end{equation}

In equation\,\ref{eq:wass_dist} $\mathcal{F}_P(\boldsymbol{X})$ and $\mathcal{F}_Q(\boldsymbol{X})$ are the cumulative distribution functions of $\boldsymbol{P}$ and $\boldsymbol{Q}$ respectively. The values for pavement-type and weather condition are on a nominal scale and use the Hellinger-distance calculated by

\begin{equation}\label{eq:avr_Hell_dist}
	\overline{\mathcal{H}}(\boldsymbol{X}_i) = \frac{1}{n}\sum_{k=1}^n \mathcal{H}(\boldsymbol{X}_{i,k}, \hat{\boldsymbol{X}}_{i,k})  \text{, where}
\end{equation}

\begin{equation}\label{eq:Hell_dist}
	\mathcal{H}(\boldsymbol{P}, \boldsymbol{Q}) = \frac{1}{\sqrt{2}} \| \sqrt{\boldsymbol{P}} - \sqrt{\boldsymbol{Q}} \|.
\end{equation}

\begin{figure*}[h]
\begin{center}
	\includegraphics{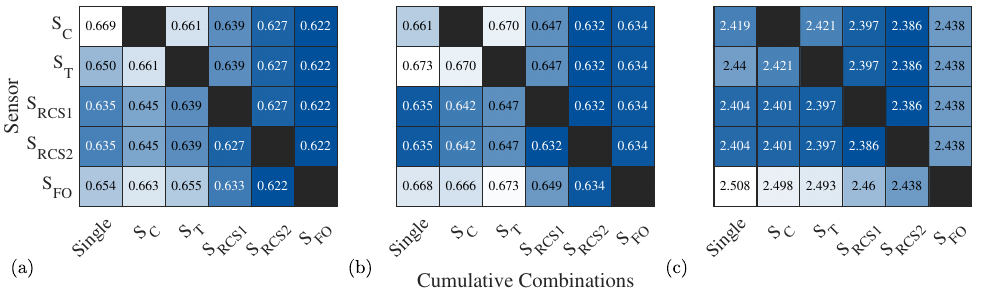}
	\caption{Average Wasserstein distance for $\boldsymbol{R}$ (a), $\boldsymbol{W}$ (b) and $\boldsymbol{\mu}_\mathrm{max}$ (c) with multiple sensor configurations}
	\label{fig:Wasser_B}
\end{center}
\end{figure*}

The result of the evaluation for the pavement type $\boldsymbol{R}$, the road weather condition $\boldsymbol{W}$ and the maximum friction coefficient $\boldsymbol{\mu}_\mathrm{max}$ is shown in Fig.\,\ref{fig:Wasser_B}. The columns represent cumulative sensor combinations with the first column being the individual sensor marked on the respective row. The other columns are the combination of the sensor marked on this column and all previous sensors. A smaller value for the respective distance measure means that over all situations the result of the inference more closely resembles $\boldsymbol{X}_{i,k}$.

It can be seen that, for all three variables, there exists a combination of sensors where the average Wasserstein-distance is lower than for all individual sensors. However, for the metric used here, all combinations of $\boldsymbol{D}$ contribute equally regardless of their probability. More possible combinations of conflicting information from the sensors can therefore result in a slightly higher average error. This can be observed for theroad weather condition $\boldsymbol{W}$ in Fig.\,\ref{fig:Wasser_B}b and the road-friction $\boldsymbol{\mu}_\mathrm{max}$ in Fig.\,\ref{fig:Wasser_B}c.

For the pavement type $\boldsymbol{R}$ the average Hellinger-distance has its smallest value of 0.662, when all sources of information are used as evidence. Among the individual sensors the RCS has the smallest average Hellinger-distance with 0.635. The camera has the highest Hellinger-distance with 0.669. This is likely due to the larger amount of possible miss-classifications of the camera.

For the road weather condition $\boldsymbol{W}$ the smallest average Hellinger-distance is achieved when combining both RCS with the camera. Since the road condition sensor $\boldsymbol{S}_{\mathrm{RCS}}$ and the camera $\boldsymbol{S}_\mathrm{C}$ directly observe the road weather condition, their average Hellinger-distance is smaller when compared to the temperature sensor $\boldsymbol{S}_\mathrm{T}$ or the friction estimator $\boldsymbol{S}_\mathrm{FO}$. The road condition sensor $\boldsymbol{S}_{\mathrm{RCS}}$ achieves the smallest average Hellinger-distance of 0.635 as a single sensor.

For the maximum friction coefficient $\boldsymbol{\mu}_\mathrm{max}$ the minimal average Wasserstein-distance of 2.386 is achieved when combining $\boldsymbol{S}_\mathrm{C}$, $\boldsymbol{S}_\mathrm{T}$ and $\boldsymbol{S}_\mathrm{RCS}$. In this case the distribution for the maximum friction coefficient $\boldsymbol{\mu}_\mathrm{max}$ is entirely calculated from a-priori knowledge. Under practical conditions this is also the combination of sensors that will be available most of the time, because the friction observer $\boldsymbol{S}_\mathrm{FO}$ only provides evidence during situations of sufficient excitation. 

\section{Experimental Results}
\begin{figure}[h]
\begin{center}
	\includegraphics{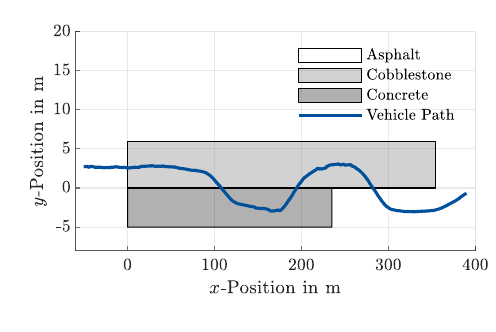}
	\caption{Path of the vehicle on the test track with wet pavement for positive $x$-values}
	\label{fig:vhlPath}
\end{center}
\end{figure}

In this section the fusion algorithm is applied to data recorded with a test vehicle. The test vehicle is a \textit{VW Golf GTE} plug-in hybrid equipped with the sensors described in Sec.\,\ref{ch:sensors}. The camera is mounted behind the wind shield and records with a resolution of 480x640\,pixels. The road condition sensors are mounted behind the two front tires. All necessary CAN-data of the vehicle is recorded with a \textit{Sirius-i} data acquisition system from \textit{Dewesoft}. The vehicles position is recorded with a differential GPS-system. The driving test was performed on a test track owned by the \textit{ZF Group} in Jeversen, Germany. During the test maneuver the vehicle drove over dry asphalt, wet cobblestone, wet concrete, and wet asphalt. The path of the vehicle is shown in Fig.~\ref{fig:vhlPath}. For positions with positive $x$-values the pavement of the test track is wet.

The camera classifies a fixed ROI in front of the vehicle which is cropped from the original image. The ROI is chosen so that its size approximately matches the footprint of the vehicle. Synchronization between camera classification and the rest of the system is done using a variable time delay $d_t$. With the distance $s$ between the upper edge of the ROI and the vehicle's front tires and the vehicles longitudinal velocity $v_t$ at time step $t$, the delay $d_t$ is calculated by 

\begin{equation}
	d_t = \frac{s}{v_t}.
\end{equation}

In the presented experimental setup, the distance $s$ is approximately \SI{6.3}{m}.

For this experiment the ground truth for the friction coefficient is unknown and allows at best an evaluation of its plausibility. Therefore, the primary metric for comparison will be calculated from the estimation of the pavement type and the road weather condition. The Ground truth is extracted from the GPS-data with respect to the conditions seen in Fig.\,\ref{fig:vhlPath}. Pavement type and road-wather condition are compared in terms of accuracy and average Hellinger-distance. The classification of the camera is split up into the corresponding classes of pavement type and road weather condition. From there the accuracy is calculated from the ratio of correct to incorrect class labels. In case the camera classifies \textit{snow}, no information about the pavement type is given and no label can be selected. These samples are given the label \textit{Nan}. For a fair comparison samples with these labels are not included in the accuracy of the pavement type for the camera and BN. The average Hellinger-distance over all samples is calculated from equation\,\ref{eq:avr_Hell_dist}. The distribution of the camera classification is calculated from the scores of the SqueezeNet's output-layer. In the case of pavement type, the score of the \textit{snow} label is evenly distributed to \textit{asphalt}, \textit{concrete} and \textit{cobblestone}. The accuracies and average Hellinger-distance are shown in Table\,\ref{tb:acc}.

\begin{table}[h]
\begin{center}
\caption{Accuracys for the classification of the pavement type $\boldsymbol{P}$ and the road weather condition $\boldsymbol{W}$} \label{tb:acc}
\begin{tabular}{lcccc}
	Algorithm &  $\mathrm{Acc}(\boldsymbol{P})$ & $\mathrm{Acc}(\boldsymbol{W})$ & $\overline{\mathcal{H}}(\boldsymbol{P})$ & $\overline{\mathcal{H}}(\boldsymbol{W})$ \\\hline
	Camera & 98.34 \% & 41.62 \% & 0.361 & 0.623 \\
	BN & 90.79 \% & 71.25 \% & 0.363 & 0.331 \\\hline
\end{tabular}
\end{center}
\end{table}

For the road weather condition $\boldsymbol{W}$ the Bayesian network achieves a higher accuracy and a lower Hellinger-distance than the standalone camera classification with a difference of \SI{29.63}{\%} and 0.292 respectively. This discrepancy mainly results from the camera, classifying the \textit{wet concrete} as \textit{snow} due to the reflection of the sky on the water surface. In these instances the cropped ROI only shows the bright reflection of the sky, which leads to a miss-classification. Some images with similar instances of miss-classifications can also be found in the SqueezeNets's training data, with the difference that here, the situation persists over a longer period of time. In the BN this is interpreted as an incorrect classification due to the information about the air temperature and the detection of water from the road condition sensor. The result is the correct classification of a \textit{wet} surface. For the pavement type the accuracy of the BN is \SI{7.55}{\%} lower than the accuracy of the camera. The difference in Hellinger-distance is only 0.002. The performance between both systems is similar, since for the BN, the main source of information about pavement type is the camera. The difference results from the connection of RCS and pavement type (compare Fig\,\ref{fig:BN_Topology}). The RCS has a slight delay in its measurement signal which results from its internal signal processing. As far as we know, this delay is not deterministic. This delay results in a lower probability for \textit{cobblestone}, when the car drives from wet concrete back to wet cobblestone at approximately \SI{200}{m}. When the camera classifies \textit{snow} while driving over the \textit{wet concrete}, the camera label did not contain direct information on the pavement type. Under these conditions the probability distribution calculated from the BN gave the labels of \textit{asphalt} and \textit{cobblestone} the highest probability which results from the cameras confusion matrix. While driving on wet cobblestone the classification of the camera was mostly correct with sporadic miss-classifications as \textit{dry cobblestone}.

\section{Conclusion}

In this article, it was shown that the robustness of the estimation of the road condition and tire friction coefficient can be increased using a Bayesian network. The Bayesian network represents a discrete probability distribution for different tire friction related parameters and enables the fusion of heterogeneous information from multiple sensor systems. It was shown with experimental data that the Bayesian network can yield higher accuracy for the classification road weather condition than an individual sensor. 

For the road pavement type only the camera could serve as a source of information, which did not improve performance. This can be addressed by adding more sources of information for the pavement type, for example map data from \cite{OpenStreetMap2017}. Additionally, the camera was most affected by the environmental conditions. Special effort should be taken to further increase the robustness of the image classification.

The information provided by the on-board temperature sensor is relatively coarse. As mentioned in Sec.\,\ref{ch:sensors}, more advanced road weather models similar to one proposed by \cite{Almkvist2023} could further enhance the results. In particular, such models can also accurately forecast the road weather class probabilities in Table 3 for a given location. Integrating this type of information into the Bayesian network will be the scope of future investigations. 

In the case presented here, the fusion is only performed for a singular time sample. With the results presented here, an extension to dynamic Bayesian networks which also consider data over a period of time could further improve the estimation.

\begin{ack}
The research for this paper was funded by the \textit{German Federal Ministry for Digital and Transport} (BMDV) under the grant number 19F2132F within the \textit{mFund} initiative. We thank Adil Murtaza Zuberi and Hauke Baumg\"artel from \textit{Hella} for their cooperation in quantifying the confusion matrix for the road condition sensor. Additionally, we thank Mirko Erich Schaper for training the SqueezeNet and preparing the required data set.
\end{ack}

\bibliography{ifacconf}             
                                                   







\appendix
\end{document}